# Analyzing the love affair of Romeo and Juliet with modern mathematical tools

Raul Isea[1*], Karl E.Lonngren[2]

[1]Institute of Advanced Studies – IDEA, Hoyo de la Puerta, Baruta, Venezuela
[2]Department of Electrical and Computer Engineering, University of Iowa, Iowa City, IA, USA

## Email address

raul.isea@gmail.com (R.I)

## Abstract

We facetiously suggest that the romance between Romeo and Juliet can be interpreted using modern terminology and include current temptations. Using this model, we consider various factors such as the time that they might spend consulting social networks, the time that they could spend alone together and along with friends as well as their tolerance of being able to waste the couple's money. The model consists of a set of differential equations which describes the relationship between them. Finally, we analyze the eigenvalues the mathematical equations in order to determine if the critical point in this model is stable or not in four different hypothetical scenarios.

## 1. Introduction

We all know the story written by William Shakespeare in 1597 entitled "Romeo and Juliet" that describes events that occurred in Verona, Italy in the 14th century. The novel tells of an impossible love affair between two rival families, the Montague's and the Capulet's. Romeo Montague and Juliet Capulet had a secret love that ended in the death of those two lovers.

This couple has been used in various scientific studies in the past [1 -3]. In the studies, there was only a consideration of the affective relationship between them. The purpose of the present work is to extend these studies to include additional factors that are found presently. These would include the additional factors such as the couple's ability to save money between them, the couple's enjoyable time possibly spent interacting with other people, and the time spent on social networks.

The narration is presented using the language of differential equations where we include the social factors that affect the relationship between them. In this preliminary model, statistics between the couple are not taken into account and the data will be analyzed in hypothetical cases using arbitrarily selected numerical values.

## 2. The Mathematical model

We represent Juliet with the symbol $J$ and Romeo with the symbol $R$. these variables are time-dependent variables. The definition for the constant $a_R$ corresponds to Juliet's feelings towards Romeo and $a_J$ corresponds to Romeo's feelings towards Juliet. The constants $c_J$ and $c_R$ reflect the capacity of Juliet responding to Romeo's feelings and Romeo responding to Juliet's feelings respectively. The parameters $m_J$ and $m_R$ correspond to the ability to save money by Juliet and Romeo respectively. We denote the toleration of the expenditure Juliet's money by Romeo with the parameter $t_J$ and Juliet's toleration of Romeo's expenditure with the parameter $t_R$. The parameters $k_J$ and $k_R$ reflect the time spent on social networks of Juliet and Romeo respectively. Two additional constants our included and they represent the mutual ability of the couple to have fun together $w$ and to save money together $s$.

$$\frac{dJ}{dt} = a_R J + c_J R + m_J J - t_J R - k_J J + sRJ + wJR$$

$$\frac{dR}{dt} = a_J R + c_R J + m_R R - t_R J - k_R R - sRJ - wJR$$

In the following, we will analyze these equations in several different cases.

## 3. Using Mathematical analysis

The solution of this system of equations uses the same methodology as explained and published before [4-7]. The system of differential equations possesses two critical points and we will discuss the equilibrium conditions for each one. Subsequently, a series of hypothetical cases derived from the present system of equations will be analyzed with selected values which help to interpret ambience between the two. In order to do this, we first need to calculate the Jacobian of the two equations which is written as:

$$\begin{pmatrix} R(s+w) + m_J - k_J + a_R & J(s+w) - t_J + c_J \\ -R(s+w) - t_R + c_R & -J(s+w) + m_R - k_R + a_J \end{pmatrix}$$

The next step is to determine the stability of the system for different initial conditions in time using arbitrarily chosen numerical values for the constants. In order to do that, the Jacobian must be evaluated at each of the critical points and the eigenvalues must be determined. The system will be stable if the value of the eigenvalue is negative.

<u>3.1 First critical point</u>.

The first critical point is $(J^*, R^*) = (0,0)$. The value of the Jacobian at this critical point is

$$J_1 = \begin{pmatrix} m_J - k_J + a_R & -t_J + c_J \\ -t_R + c_R & m_R - k_R + a_J \end{pmatrix}$$

where the subscript 1 represent the first critical point. There are two eigenvalues that are found to be:

$$\frac{\mp\sqrt{(a_J - a_R + k_J - m_J - k_R - m_R)^2 + 4(c_J - t_J)(c_R - t_R)}}{2} \mp$$

$$\mp \frac{(m_J - k_J + a_R) \mp (m_R - k_R + a_J)}{2}$$

These two values will be real if convention prescribed by

$$(a_J - a_R + k_J - m_J - k_R - m_R)^2 > 4(c_J - t_J)(c_R - t_R)$$

This critical point is stable when the eigenvalue is negative and equal to

$$-\frac{1}{2}\sqrt{(a_J - a_R + k_J - m_J - k_R - m_R)^2 + 4(c_J - t_J)(c_R - t_R)} -$$

$$-\frac{1}{2}(m_J - k_J + a_R) - (m_R - k_R + a_J)$$

3.1.1 Case 1:
Romeo and Juliet are in love with each other which indicates the constants should assume the values ($a_R = 1, a_J = 1$) and ($c_R = 1, c_J = 1$). In addition, they do not spend time on social networks ($k_R = 0, k_J = 0$) and have a lot of fun together ($w = 1$). In addition, they do not care if their partner squanders/spends their money without any limitation ($t_R = 0, t_J = 0$) and they save money together ($m_R = 0.5, m_J = 0.5$). In this case, the eigenvalue is positive and equal to 0.5 which indicates that the relationship is not stable.

3.1.2 Case 2:
Romeo is madly in love, but Juliet does not respond to this love with the same ardor. The constants will be chosen to be ($a_R = 1, a_J = 0.5, c_R = 1, c_J = 0.5$). They spend a considerable amount of time on social networks in order to keep track of the other persons activities. ($k_R = 1, k_J = 1$). Juliet does not spend much time with Romeo ($w = 0.3$). As far as spending money, Romeo spends a different amount than Juliet and the constants I chose to do have the values ($t_R = 1, t_J = 0.2, m_R = 1, m_J = 0.1, s = 0.7$). In this hypothetical scenario and at the critical point, the eigenvalues are not stable (value equal to + 0.1). A similar result would ensue if the conditions had been reversed.

### 3.1.3 Case 3:

Romeo is madly in love with Juliet but Juliet does not reciprocate this love $(a_R = 1, a_J = 0, c_R = 1, c_J = 0)$. They spend a lot of time on social networks$(k_R = 1, k_J = 1)$ and Juliet does not spend much time with Romeo$(w = 0)$.Regarding money, Romeo spends a lot of money but he does not do it in the same fashion as Juliet$(t_R = 0.2, t_J = 0.2, m_R = 0.1, m_J = 0.1.$This hypothetical scenario is not possible because the eigenvalue has an imaginary value (0.05+0.40i).

### 3.1.4 Case 4:

Romeo loves Juliet but Juliet only likes Romeo leading to the chosen parameters $(a_R = .6, a_J = 0.1, c_R = .3, c_J = 0.1)$. Romeo tries to follow Juliet using the social network $(k_R = 1, k_J = 0.4)$. They have some time together $(w = 0.4)$.Romeo spends money in the same manner as Juliet$(t_R = 0.5, t_J = 0.5, m_R = 0.5, m_J = 0.5)$. This particular scenario these to a stable situation since the eigenvalue is negative and has the value of -0.47

### 3.2 Second critical point

The second critical point is $(J^*, R^*)$:

$$J^* = (t_J - c_J)(t_R - c_R) + (k_J - a_R - m_J)(m_R - k_R + a_J)$$
$$R^* = -(t_J - c_J)(t_R - c_R) - (k_J - a_R - m_J)(m_R - k_R + a_J)$$

The expression of the Jacobian is extremely complicated and the eigenvalues for this case are obtained from the following expression:

$$\frac{-\sqrt{(B-A)^2 + 4\,CD} - A - B}{2}$$

where we have defined

$$A \equiv -\frac{(t_J - c_J)(t_R + k_J - m_J - c_R - a_R)}{k_R - a_J - c_J - m_R + t_J}$$

$$B \equiv -\frac{(t_J - c_J - m_R + k_R - a_J)(t_R - c_R)}{k_J - a_R - c_R - m_J + t_R}$$

$$C \equiv \frac{(t_J - c_J - m_R - a_J + k_R)(-k_J + m_J + a_R)}{k_J - a_R - c_R - m_J + t_R}$$

$$D \equiv \frac{(m_R - k_R + a_J)(t_R - c_R - m_J + k_J - a_R)}{k_R - a_J - c_J - m_R + t_J}$$

Finally, we analyzed the same four cases analyze previously. Using the same conditions as in the first critical point, we will only indicate the numerical result.

### 3.2.1 Case 1:
The critical point in this case is $(R^*, J^*) = (-0.42, 0.42)$ and the eigenvalue is equal to-0.5. This means that there is a stable relationship.

### 3.2.2 Case 2:
In this hypothetical scenario, the critical point is $(R^*, J^*) = (-0.8, 0.62)$ and the eigenvalue is -0.21. This means that there is a stable relationship.

### 3.2.3 Case 3:
This scenario $(R^*, J^*) = (1.6, 0.4)$ is possible because the eigenvalue is -0.17. This means that there is a stable relationship.

### 3.2.4 Case 4:
In this hypothetical scenario $(R^*, J^*) = (-0.5, -0.8)$ is not stable because the eigenvalue is complex positive value (0.28-0.52i).

## 4. Conclusion

In this somewhat facetious analysis using modern mathematical techniques and terminology, we have analyzed the stability of the relationship between Romeo and Juliet. The choice for the numerical values for all of the parameters was entirely arbitrary. With the rapid development of computer technology and numerical techniques, we believe that a more comprehensive analysis can and will be performed in order to bring numerical technology into different venues. Finally, there are other factors that may help one to understand the relationship between couples that should be considered. It may be necessary to correlate them with data from interviews and follow-ups between couples over an extended period of time. It only remains to be seen what are the ideal conditions that lead to a romance with a total dedication to each other.

## Acknowledgements

This manuscript is dedicated to our wives who have loved and supported us in our endeavours throughout our lives.